\documentclass[letter]{aa}  

\usepackage{graphicx}
\usepackage[varg]{txfonts}
\usepackage{longtable}
\usepackage{lscape}
\usepackage[]{hyperref}

\newcommand{\teff}{\ensuremath{T_{\mathrm{eff}}}}
\newcommand{\logg}{\ensuremath{\log g}}
\newcommand{\vmicro}{\ensuremath{\xi_\mathrm{micro}}}
\newcommand{\feh}{\ensuremath{\left[\mathrm{Fe}/\mathrm{H}\right]}}

\newcommand{\ATLAS}{{\tt ATLAS9}}

\newcommand{\MULTI}{{\tt MULTI}}
\newcommand{\SYNTHV}{{\tt SynthV}}
\newcommand{\TURBO}{{\tt TURBOSPECTRUM}}

\begin{document}

\title{Abundance of beryllium in the Sun and stars:\\
The role of non-local thermodynamic equilibrium effects}

\author
        {
        S.~Korotin\inst{1}
        \and
        A.~Ku\v{c}inskas\inst{2}
        }

\institute
        {
        Crimean Astrophysical Observatory, Nauchny 298409, Crimea
        \and
        Institute of Theoretical Physics and Astronomy, Vilnius University, Saul\.{e}tekio al. 3, Vilnius, LT-10257, Lithuania
        }

\date{Received ; accepted}

 
\abstract
{
Earlier studies have suggested that deviations from the local 
thermodynamic equilibrium (LTE) play a minor role in the formation of 
\ion{Be}{ii} 313 \,nm resonance lines in solar and stellar atmospheres. 
Recent improvements in the atomic data 
allow a more 
complete model atom of Be to be constructed and the validity of these claims 
to be reassessed using more up-to-date atomic physics.}
{The main goal of this study therefore is to refocus on the role of non-local 
thermodynamic equilibrium (NLTE) effects in the formation of \ion{Be}{ii} 
313.04 and 313.11\,nm resonance lines in solar and stellar atmospheres.}
{For this, we constructed a model atom of Be using new atomic data 
that recently became available. The model atom contains 98 levels and 
383 radiative transitions of \ion{Be}{i} and \ion{Be}{ii} and uses the most up-to-date
collision rates with electrons and hydrogen. This makes it the most complete 
model atom used to determine 1D~NLTE 
solar Be abundance and to study the role of NLTE effects in the 
formation of \ion{Be}{ii} 313\,nm resonance lines.}
{We find that deviations from LTE have a significant influence on the strength of the \ion{Be}{ii} 
313\,nm line in solar and stellar atmospheres. For the 
Sun, we obtained the 1D~NLTE Be abundance of ${\rm A(Be)_{\rm NLTE}}=1.32\pm 0.05$, which is in 
excellent agreement with the meteoritic value of ${\rm A(Be)}=1.31\pm 0.04$. 
Importantly, we find that NLTE effects become significant in FGK stars. 
Moreover, there is a pronounced variation in 1D~NLTE--LTE abundance 
corrections with the effective temperature and metallicity. Therefore, 
contrary to our previous understanding, the obtained results indicate that 
NLTE effects play an important role in Be line formation in 
stellar atmospheres and have to be properly taken into account in 
Be abundance studies, especially in metal-poor stars.}
{}


\authorrunning{Korotin \& Ku\v{c}inskas}
\titlerunning{Beryllium abundance in the Sun and stars}

\maketitle

\section{Introduction}

Studies of Be abundance are of interest in various astrophysical  contexts. 
Because Be is synthesised mostly via the cosmic ray spallation in the 
interstellar medium \citep[e.g.][]{SZP21} and is quickly destroyed in stars 
(but see \citealt{Maia2015}), 
Be may serve as a useful tracer of mixing in stellar interiors 
\citep[e.g.][]{SPC10}. In the context of Galactic chemical evolution, Be may 
provide valuable information about the composition and role of cosmic rays in 
the chemical enrichment of Galactic interstellar matter (e.g. \citealt{P12}).

The only Be abundance indicators in the optical range are \ion{Be}{ii} resonance 
lines located at 313.04 and 313.11\.nm. Earlier studies have shown that non-local thermodynamic equilibrium (NLTE) 
effects lead to strong over-ionisation of \ion{Be}{i} in the solar atmosphere, 
with a significant increase in \ion{Be}{ii} concentration in the \ion{Be}{ii} 
313\,nm line-forming region \citep{Asplund05}. This is because the strongest 
lines and the ionisation cutoff of \ion{Be}{ii} are located in the UV, where solar 
radiative flux exceeds that of the Planck function; this leads to 
the over-ionisation of \ion{Be}{i}. For the same reason, the higher levels of 
\ion{Be}{ii} become overpopulated and the lowest levels underpopulated. 
Thus, the first effect makes \ion{Be}{ii} lines appear stronger 
and the second weaker. A number of studies performed until now have suggested 
that the two processes effectively cancel each other out and lead to a zero 
NLTE--LTE (local thermodynamic equilibrium) Be abundance correction for the Sun \citep{Garcia95,Takeda11}. With this, 
the `standard' solar photospheric Be abundance, ${\rm A(Be)}=1.38\pm0.09$ 
\citep{Asplund21}, is slightly higher than that measured in meteorites, 
${\rm A(Be)}=1.31\pm0.04$  \citep{Lodders21}.

Amongst the most important advances in Be abundance studies, new atomic data 
recently became available. In this context, collisional 
cross-sections with electrons and hydrogen obtained from quantum mechanical 
computations by \citet{Dipti19} and \citep{Yakovleva16} are of particular interest as 
these effects play a crucial role in NLTE line formation. These 
atomic data were not yet available to earlier studies of Be NLTE line formation 
in solar and stellar atmospheres, and thus collisions were treated in a 
simplified fashion. Therefore, the
availability of new atomic data on the one hand and the discrepancy between 
the photospheric and meteoritic solar Be abundances on the other made it 
desirable to re-determine solar Be abundance and to reassess the role of 
NLTE effects in Be spectral line formation, both in the Sun and other types 
of stars.

The motivation behind the present study was therefore threefold. 
First, using the 
most-up-to-date atomic data we constructed a new model atom of Be to 
assess the role of NLTE effects on the \ion{Be}{ii} 313\,nm  
line formation in solar and stellar atmospheres. Second, the new model atom 
was used to obtain an updated value of 1D~NLTE solar Be abundance. And finally, 
we investigated the role of NLTE effects on Be line formation in 
stellar atmospheres with a goal of providing a grid of 1D~NLTE--LTE Be 
abundance corrections that could be used for Be abundance studies in different 
types of stars.

\section{Methodology of 1D~NLTE Be abundance analysis}

\subsection{Observational data\label{sect:obs-data}}

The only Be lines available for stellar abundance analysis in the optical 
range are those of \ion{Be}{ii} located at 313\,nm. For the calibration 
of solar flux in this spectral range (Sect.~\ref{sect:missing-UV-opacity}) 
we a used solar irradiance spectrum obtained with the SOLar SPECtrometer 
(SOLSPEC) instrument of the SOLAR payload on board the International Space 
Station \citep{Meftah18}. The SOLSPEC spectrum covers a 
range of $\approx165-3000$\,nm with a resolution of $\approx0.1-0.2$\,nm 
in the UV.

For the calibration of the solar flux in the UV--blue spectral range we 
also used two additional low-resolution solar irradiance flux spectra: 
(a) the Upper Atmosphere Research Satellite (UARS) irradiance spectrum 
obtained in the range of $115.0-420.0$\,nm with a resolution 
of $\approx1$\,nm \citep{Woods96}; and (b) the 
Compact Spectral Irradiance Monitor (CSIM) irradiance spectrum in the range 
of $200.0-2600.0$\,nm at $\approx2$\,nm resolution\footnote{https://lasp.colorado.edu/home/csim/.}.

The solar flux atlas from \citet{Kurucz84} was used for solar Be abundance 
determination (Sect.~\ref{sect:solar-Be-abundance}). It covers a range of 
$296-1300 $\,nm and has a resolution of $\approx350000$ in the near UV.

\subsection{Model atom of Be}


Having the first ionisation potential of 9.32\,eV, Be in the atmospheres of FGK stars
is available in the neutral and single-ionised states; thus, both stages should be 
taken into account for computing NLTE level departure coefficients. Our Be 
model atom therefore consists of 48 levels of \ion{Be}{i} ($n\leq8$, $l\leq4$), 49 levels 
of \ion{Be}{ii} ($n\leq12$, $l\leq5$), and a ground level of \ion{Be}{iii}. 
Level energies were taken from the NIST database, \citet{Kramida12}, and level fine structure was not 
taken into account. The lowest ionisation energy for \ion{Be}{i} (from level 
8d~3D) of 0.22\,eV ensures a reliable connection between the \ion{Be}{i} 
and \ion{Be}{ii} stages of the model atom at $T\gtrsim2000$\,K. Ionisation 
energy from the highest level in the model atom of \ion{Be}{ii} is slightly higher, 
0.38\,eV ($T\gtrsim4000$\,K), but this is not 
critical, because the fraction of \ion{Be}{iii} in the atmospheres of cool 
stars is very small. 

For the computation of departure coefficients, 230 bound--bound radiative 
transitions were taken into account for \ion{Be}{i} and 153 for \ion{Be}{ii}. 
We used oscillator strengths from the TOPBase \citep{Cunto93}. 
Importantly, the majority of Be lines are very weak, and thus uncertainties in 
the oscillator strengths play a negligible role. For the strongest transitions 
that occur from the lowest two levels of \ion{Be}{i}, oscillator strengths 
were taken from \citet{Tachiev99}, while for \ion{Be}{ii} 313\,nm resonance 
lines we used quantum mechanical computation data from \citet{Yan98}. 
Photoionisation cross-sections were taken from the TOPBase for 
all \ion{Be}{i} and \ion{Be}{ii} levels except for levels f and g 
of \ion{Be}{ii}, in which case a hydrogen-like approximation was used. 
Our tests have shown that one order variation in the cross-sections
of levels f and g of \ion{Be}{ii} had a 
negligible effect on the population numbers of the lowest \ion{Be}{ii} levels.

Collision rates for the transitions between 
the 19 lowest levels of \ion{Be}{i} were taken from \citet{Dipti19} and those
for transitions between the 10 higher-lying 
levels (up to and including 6s~1S) from the ADAS database \citep{Summers11}.
The ADAS collisional rates were also used to account for the transitions 
between the 14 lowest levels of \ion{Be}{ii}.
For the rest of the bound–bound collisional transitions, we adopted the 
formulae from \citet{Regemorter62} for permitted transitions and those from
\citet{Allen73} for the forbidden ones. Collisional ionisation rates
were calculated using the formula from \citet{Seaton62}.

To account for inelastic transitions with hydrogen, we used quantum mechanical 
data from \citet{Yakovleva16} for the first 14 levels of \ion{Be}{i}. For the 
remaining levels, Drawin's approximation \citep{Drawin68} was used in the form 
provided by \citet{Steenbock84}, 
with S$_{\rm H}$=0.1. Variation in S$_{\rm H}$ values from 0 to 1 had no 
effect on the line strength in the spectra of dwarf and giant stars with 
$3500<\teff<6500$\,K, with changes in the line equivalent width of 
$<0.1\%$. Once again, transitions from the lowest levels of \ion{Be}{i} play 
a major role, and for these cases quantum mechanical data were used.

\subsection{Spectral line synthesis\label{sect:line-synthesis}}

The \ATLAS\ model atmospheres that were used in spectral line synthesis 
calculations were computed using opacity distribution functions (ODFs)  from 
\citet{Mezaros12}. The ODFs were constructed using updated absorption lists of 
molecular lines, which is of critical importance for calculating the atmospheres 
of cool stars.
For the calculation of synthetic spectra in the vicinity of \ion{Be}{ii} lines,
level departure coefficients were computed using the \MULTI\ code 
\citep{Carlsson86} in the form modified by \citet{Korotin99} 
to enable the use of \ATLAS\ opacities. 
Kurucz's ODFs \citep{Mezaros12} were used to account for 
line blanketing only when computing photoionisation rates.
Synthetic spectra were produced using the \SYNTHV\ code \citep{Tsymbal19}, with 
\ion{Be}{ii} line profiles computed in NLTE (using the departure coefficients 
obtained with \MULTI) and those of other lines calculated in LTE. 
The atomic line data were taken from the VALD3 database \citep{Ryabchikova15}. 
We used data from \citet{Barklem2000} to account for the van der Waals broadening of \ion{Be}{ii} lines. 
For the Sun, we also used the \TURBO\ code \citet{Plez12}
(v.19; Sects.~\ref{sect:missing-UV-opacity} and 
\ref{sect:solar-Be-abundance}).

\section{NLTE effects in Be spectral line formation and Be abundances in the 
Sun and stars}

\subsection{Missing near-UV opacity in the solar spectrum\label{sect:missing-UV-opacity}}

Determination of the continuum in the vicinity of \ion{Be}{ii} 313\,nm 
resonance lines is complicated in stars of higher metallicity because this 
spectral range is heavily blended with various atomic and molecular lines. 
Inevitably, this takes a toll on the accuracy of determined Be abundances.


In addition, the accuracy of Be abundances is greatly influenced by the 
reliability of the continuous opacity that is used in spectral synthesis 
calculations. For example, \citet{Balachandran98} concluded that it was 
necessary to increase the continuum opacity by 1.6 times in the synthetic 
spectrum around 313\.nm to obtain a reasonable fit with the observed solar flux.
\citet{Fontenla15} showed that taking absorption from the 
dissociation of CH, NH, and OH into account may also affect the continuous opacity 
noticeably. However, \citet{Carlberg18} (hereafter C18) demonstrated 
that even for cool stars the additional contribution from molecules 
does not exceed 4\%. Therefore, in order to reconcile the synthetic spectrum 
of the Sun with the observed spectrum, an additional factor of 1.2 was introduced 
in C18 to increase the total opacity in the continuum. \citet{Bell01} 
have concluded that an adequate description of the UV flux is obtained if 
the opacity due to  b--f transitions of \ion{Fe}{i} is added and increased by a 
factor of two.
This conclusion was supported by \citet{Asplund04}, who showed that a 
50\% increase in the opacity was needed to obtain O abundances 
from OH lines near 313\,nm that are consistent with those determined using 
OH lines in the infrared.


To check the effects of missing opacity in our spectral synthesis calculations, 
we computed solar flux in the range of $250-400$\,nm  with a step of $0.005$\,nm 
using the \SYNTHV\ and \TURBO\ spectral synthesis packages. Both codes used 
identical Fe b--f opacities taken from the TOPBase \citep{Cunto93}, and synthesis of 
all spectral lines was performed under the assumption of LTE. In addition, a block of opacities resulting from the 
dissociation of CH and NH molecules was added to the calculations, as proposed in C18. For comparison, 
we used the SOLSPEC, CSIM, and UARS solar irradiance spectra 
(Sect.~\ref{sect:obs-data}) converted to flux on the 
solar surface. Spectral synthesis was performed 
using the solar \ATLAS\ model computed by \citet{Castelli03}
\footnote{https://wwwuser.oats.inaf.it/castelli/sun/ap00t5777g44377k1asp.dat}, 
with the line data taken from the VALD3 database \citep{Ryabchikova15}, solar chemical composition from \citet{Asplund21}, and 
a microturbulence velocity of 1\,km/s. 
For the comparison with observations, the resolution of the synthetic flux was 
degraded to that of the observed spectra.

\begin{figure}[tb]
\resizebox{\hsize}{!}{\includegraphics{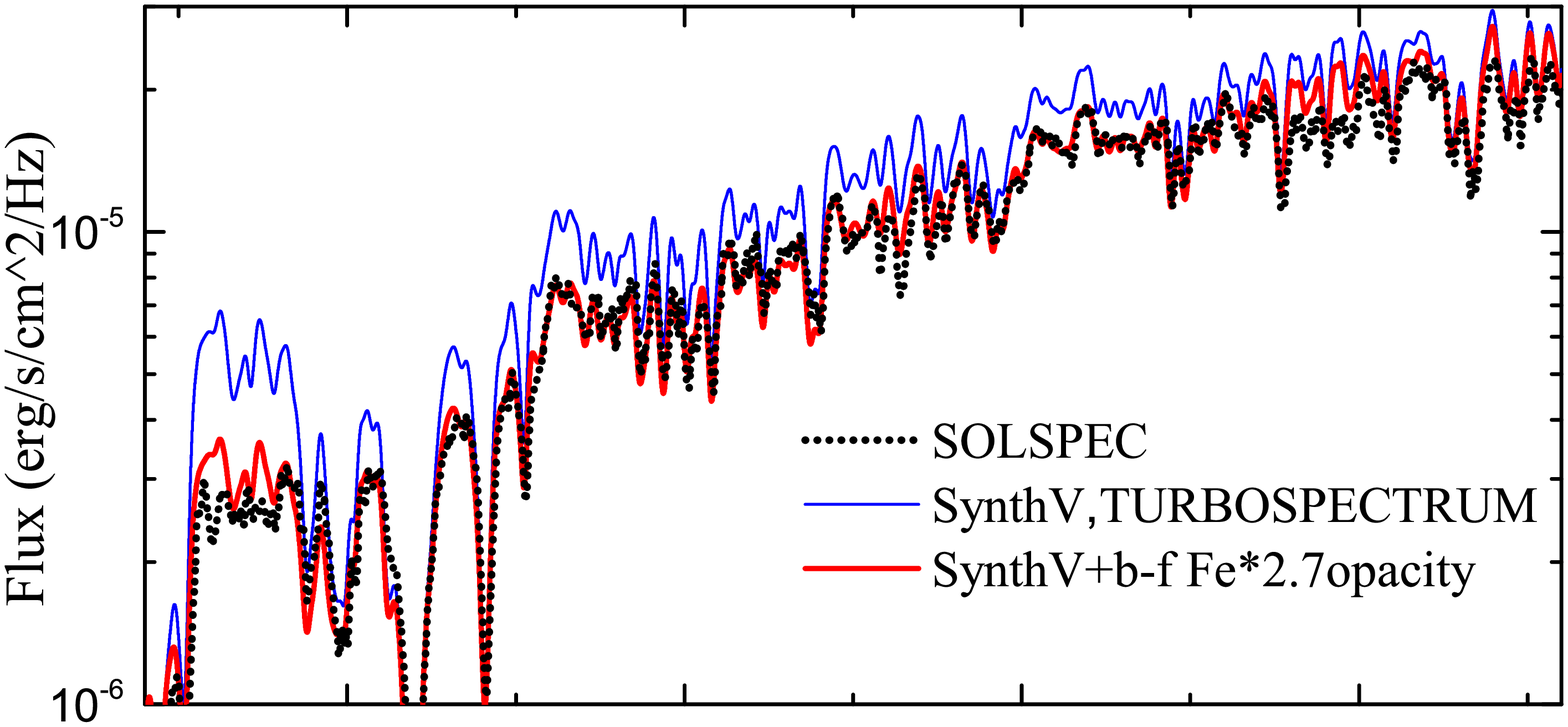}}
\resizebox{\hsize}{!}{\includegraphics{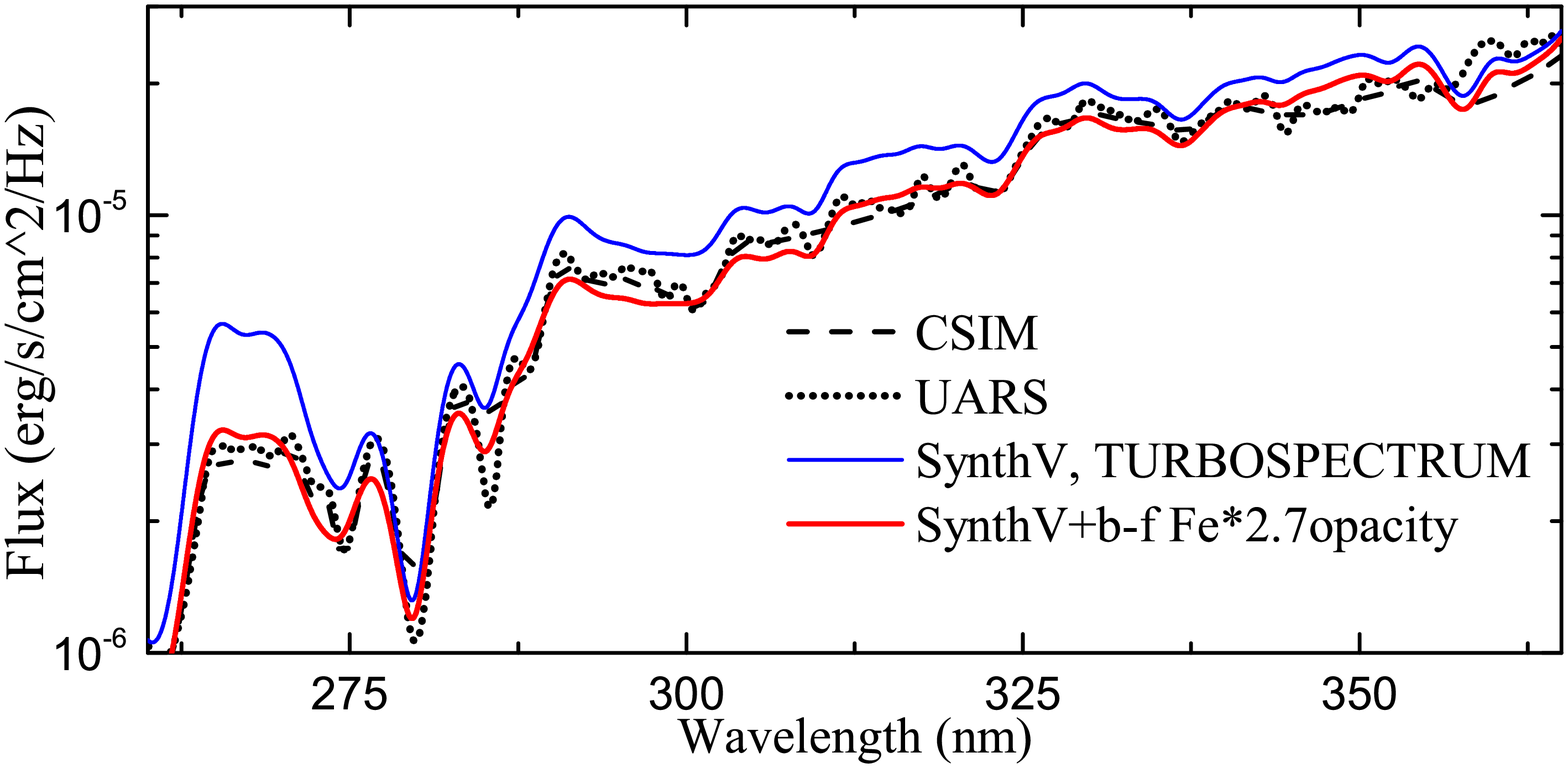}}
\caption
{Synthetic and observed solar fluxes in the blue--UV spectral range. 
Top: Comparison of synthetic fluxes with  the observed SOLSPEC flux. 
Bottom: Comparison of synthetic and observed CSIM and UARS fluxes.
}
\label{fig:solar-flux}
\end{figure}

The results of our computations demonstrate that the solar flux in the vicinity 
of Be lines predicted by spectral synthesis with the \SYNTHV\ and \TURBO\ 
packages is nearly identical and, at the same time, is $25-30\%$ higher than observed (Fig.~\ref{fig:solar-flux}). 
On the other hand, there is an excellent agreement 
between the observed and synthetic spectra at $\gtrsim390$\,nm, where the 
differences do not exceed $1-2\%$.

To bring synthetic and observed fluxes of the Sun into agreement  
in the UV region, we artificially increased the b--f opacities of Fe by a 
factor of 2.7 at $>265.0$\,nm. Such an approach was adopted earlier by 
\citet{Bell01}, in which case the Fe opacities were increased by a factor of two. 
With this adjustment, we obtained a very good agreement between 
the synthetic fluxes computed with the \SYNTHV\ and \TURBO\ codes and the 
observed SOLSPEC flux (Fig.~\ref{fig:solar-flux}, top panel). 
The adjusted synthetic spectra also agree reasonably well with the observed 
low-resolution CSIM and UARS fluxes (Fig.~\ref{fig:solar-flux}, bottom panel).

\subsection{Modifications to the VALD3 line list\label{sect:modifications-line-list}}

As indicated in Sect.~\ref{sect:line-synthesis}, we used atomic line data from 
the VALD3 database \citep{Ryabchikova15} to determine the solar Be abundance.  
We made several further improvements related to (a) the lines that directly affect the Be abundance 
determination; and (b) the lines that do not affect Be abundances but give a 
better fit in the immediate neighbourhood of \ion{Be}{ii} lines. For this we 
roughly followed the prescription used by \citet{Takeda11}, 
\citet{Delgado12}, and C18 in their solar Be abundance studies. 
For group (a) lines, we added the \ion{CH}{} 313.0367\,nm line from 
\citet{Delgado12}; increased the $\log gf$ from $-2.222$ to $-1.561$ for 
the \ion{Ti}{i} 313.0377 line, as suggested by \citet{Takeda11} and 
\citet{Delgado12}; and shifted the wavelength of the 
\ion{Mn}{i} 313.1036\,nm line by $-0.002$\,nm and increased its 
$\log gf$ to $-0.550$, following the recommendation of 
\citet{Ashwell05}. We also made adjustments 
to several group (b) lines: removed the \ion{Fe}{i} 312.9619\,nm line, 
as recommended by C18; increased the $\log gf$ from $-0.401$ to $-0.201$ for 
the \ion{Cr}{i} 313.1216\,nm line; and added \ion{CH}{} 313.0644\,nm and 
\ion{OH}{} 313.1366\,nm lines from the list of \citet{Delgado12}. 
We did not add any `hypothetical' lines, such as \ion{the Ti}{ii} 313.1039\,nm 
line introduced by C18 or  the \ion{Fe}{i} 313.1043\,nm line by \citet{Takeda11}. 
The modified line list is provided in Appendix~\ref{app-sect:line-list}.

\begin{figure}[tb]
\resizebox{\hsize}{!}{\includegraphics{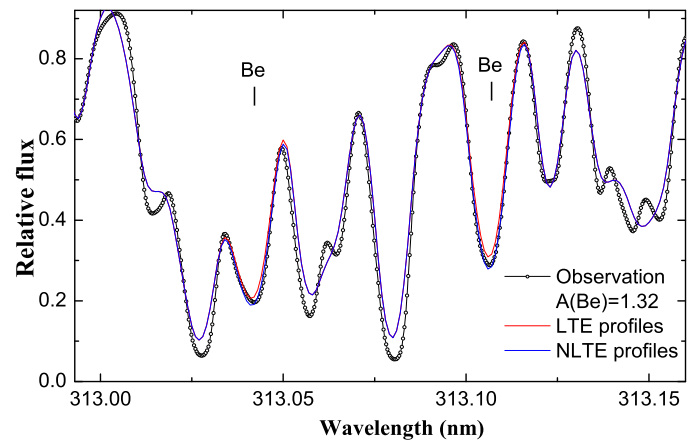}}
\caption
{Observed solar spectrum from the Kurucz atlas (dots) and the best-fitted 
synthetic \SYNTHV\ spectra (solid lines) in the vicinity of \ion{Be}{ii} lines. 
The solid blue line denotes a synthetic spectrum with the \ion{Be}{ii} lines 
computed in NLTE and the remaining lines in LTE. Solid red lines are 
\ion{Be}{ii} profiles computed in LTE with the same Be abundance. 
}
\label{fig:solar-Beii-fit}
\end{figure}

The adjustments to the line list described above and the modification of the continuum 
opacity (Sect.~\ref{sect:missing-UV-opacity}) allowed us to obtain an adequate 
fit of the solar spectrum in the vicinity of the \ion{Be}{ii} lines 
(Fig.~\ref{fig:solar-Beii-fit}).

\subsection{NLTE effects and Be abundance in the Sun\label{sect:solar-Be-abundance}}

\begin{figure}[tb]
\centering
\includegraphics[width=8cm]{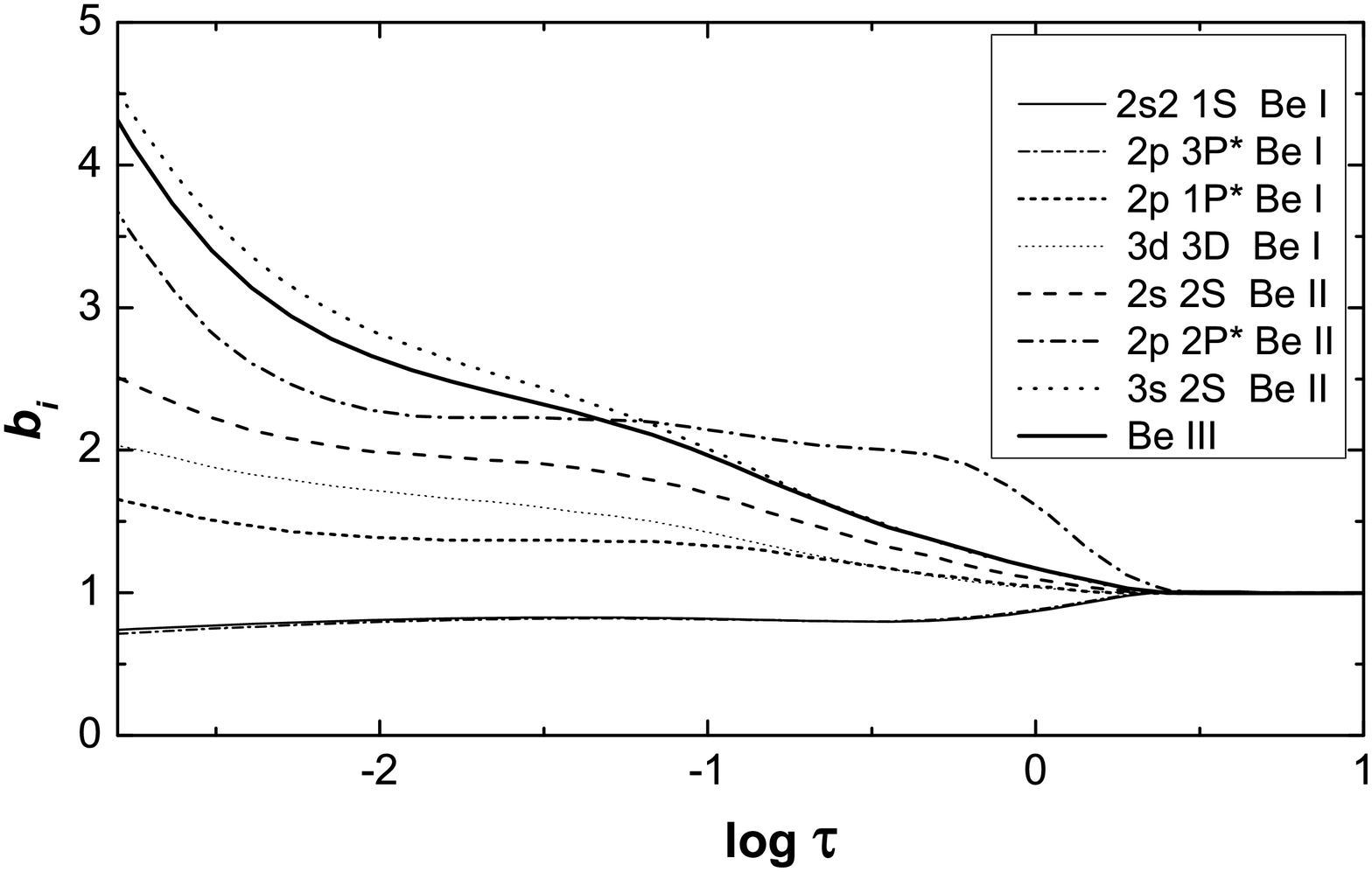}
\caption
{Departure coefficients, $b_{\rm i}$, of different atomic levels of \ion{Be}{i} 
and \ion{Be}{ii} in the solar atmosphere, plotted as a function of optical 
depth.
}
\label{fig:solar-b-factors}
\end{figure}

Using the solar atlas from \citet{Kurucz84} together with synthetic 
spectra computed using the \SYNTHV\ code and departure coefficients obtained 
with the \MULTI\ code, we determined a solar Be NLTE abundance of 
${\rm A(Be)_{\rm NLTE}}=1.32\pm 0.05$. This value 
is slightly  lower than the ${\rm A(Be)_{\rm NLTE}}=1.38\pm0.09$ obtained in \citet{Asplund21} but is in excellent agreement with the 
meteoritic value of ${\rm A(Be)}=1.31\pm0.04$ \citep{Lodders21}. 
We also repeated the analysis using the \citet{Holweger1974} semi-empirical
solar model atmosphere. In this case we obtained ${\rm A(Be)_{\rm NLTE}}=1.35\pm 0.05$
(we note that the NLTE--LTE abundance corrections determined with the \ATLAS\ 
and \citet{Holweger1974} models were identical, but the \ion{Be}{ii} line was slightly weaker 
in the latter case, thus leading to a slightly higher Be abundance).

Similar to what was found earlier, we find that in the solar 
atmosphere the lowest levels of \ion{Be}{i} are strongly underpopulated, while 
those of \ion{Be}{ii} are overpopulated (Fig. \ref{fig:solar-b-factors}). However, overpopulation of 
the higher levels is stronger than that of the ground level. 
This subtle effect leads to a slight strengthening of \ion{Be}{ii} resonance 
lines at 313.04 and 313.11\,nm, with their equivalent widths becoming higher 
by 5\% and 7\%, respectively. Therefore, importantly -- and contrary to the 
findings of earlier studies \citep[e.g.][]{Garcia95,Takeda11} -- our results suggest 
that NLTE effects may strengthen the \ion{Be}{ii} 313.04 
and 313.11\,nm resonance lines significantly, 
leading to a total solar Be 
NLTE--LTE abundance correction of $-0.07$\,dex. 

Several factors may account for this difference with the earlier studies. 
First, the \ion{Be}{i} and \ion{Be}{ii} stages of the Be model atom used in our 
study contain more atomic levels and thus allow the 
b--b and b--f transitions that involve the highest levels to be better accounted for. Second, we used 
the most up-to-date collisional rates with electrons and hydrogen 
obtained using quantum mechanical computations. 
In fact, quantum mechanical cross-sections of collisional ionisation by 
hydrogen from the first two levels of \ion{Be}{i} obtained by 
\citet{Yakovleva16} are two to five orders of magnitude lower than those predicted 
in the Drawin's approximation. At the same time, the cross-sections of 
\citet{Yakovleva16} are two to four orders of magnitude higher than in the 
Drawin's approximation for the higher levels of \ion{Be}{i}. 
Inevitably, this should lead to changes in the ionisation balance 
between \ion{Be}{i} and \ion{Be}{ii}. 

To verify the importance of the factors described above, we used 
the same 98-level model atom but with the collisional rates computed using 
the Drawin's approximation (${\rm S}_{\rm H}=1$). With this, we obtained 
an NLTE--LTE Be abundance correction of --0.035\,dex. If the model atom is 
simplified further to consist of ten levels of \ion{Be}{i}, five levels of \ion{Be}{ii}, 
the ground level of \ion{Be}{iii}, and collisional rates with electrons and 
hydrogen computed using van Regemorter's and Drawin's formulae, respectively, we obtain the abundance correction
of $-0.01$\,dex, which is in excellent agreement with those determined by 
\citet{Garcia95} and \citet{Takeda11}. Repeating calculations with the 16-level
model atom at several other combinations of \teff\ and $\log g$, we obtain 
abundance corrections that agree very well with those of \citet{Takeda11} (Table \ref{app-tab:NLTE-corr-comparison}).
This leads us to the conclusion that the 
differences between the solar Be abundance obtained by us and other authors are due 
to the increased complexity and realism of our model atom.


Clearly, it is always desirable to check whether the NLTE abundances of a given 
element agree if they are determined from lines that form due to 
transitions from different multiplets in different ionisation stages. This, 
however, is not possible in the case of Be because the lines of \ion{Be}{i} 
are located in the UV (234.8, 249.4, 265.0\,nm) and are very heavily 
blended in the solar spectrum.

Finally, we recall that the missing UV opacity (Sect.~\ref{sect:missing-UV-opacity}) 
has a direct influence on the determined Be abundance. For example, using 
standard (i.e. uncorrected) \ATLAS\ opacities and our new Be model atom, we 
obtain the solar Be abundance of ${\rm A(Be)_{\rm NLTE}}=1.09\pm0.05$. 
By adding b--f opacities of \ion{Fe}{i} and \ion{Fe}{ii} from the \TURBO\ 
package, as well as opacities resulting from the dissociation of CN and NH (C18), 
one obtains ${\rm A(Be)_{\rm NLTE}}=1.18\pm0.05$. This value is 
0.14\,dex lower than that obtained by us with the adjusted Fe b--f 
opacities.
One may anticipate that 
this effect will be important in other types of stars as well and thus has to 
be properly taken into account.

\subsection{NLTE effects and Be abundances in stellar atmospheres\label{sect:NLTE-corr}}

\begin{figure*}[tb]
\centering
\includegraphics[width=18.5cm]{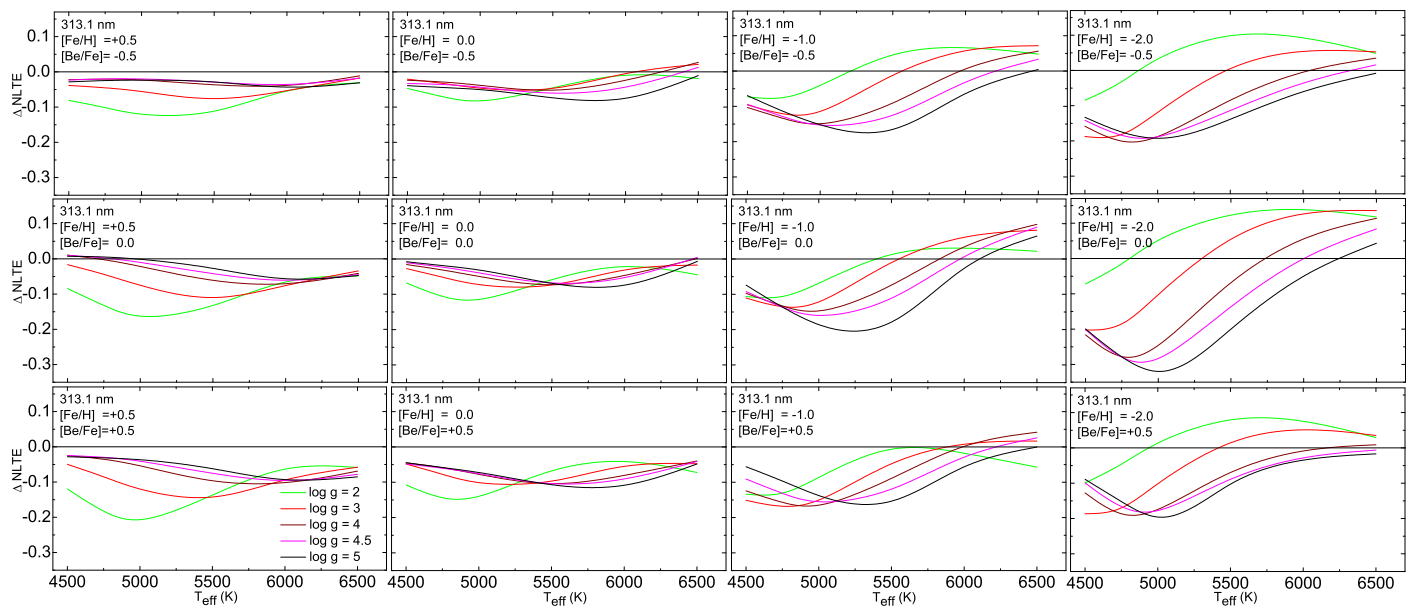}
\caption
{1D~NLTE--LTE Be abundance corrections for the \ion{Be}{ii} 313.11\,nm resonance 
line, for different values of \teff, $\log g$, \feh, and ${\rm [Be/Fe]}$.
}
\label{fig:nlte-be-corr}
\end{figure*}


To estimate the influence of NLTE effects on the formation of \ion{Be}{ii} 
313.04 and 313.11\,nm resonance lines in the atmospheres of other types of 
stars, we computed a small grid of 1D~NLTE--LTE Be abundance corrections. For this 
we used \ATLAS\ model atmospheres with $4500\leq\teff\leq6500$\,K, 
$2.0\leq\log g\leq5.0$, $-2.0\leq\feh\leq+0.5$, $\vmicro=2$\,km/s, and 
${\rm [Be/Fe]}=-0.5; 0.0; +0.5$. Abundance corrections are provided in Fig.~\ref{fig:nlte-be-corr}
and Table~\ref{app-tab:NLTE_corrections}.

Importantly, the obtained corrections show that NLTE effects may be substantial 
in other types of stars as well (Fig.~\ref{fig:nlte-be-corr}). In extreme 
cases they may reach up to --0.3\,dex, with a general tendency of corrections being larger at 
the lower effective temperatures and higher gravities. The correction may be 
positive or negative depending on which process -- over-ionisation or 
over-excitation -- dominates. There is also a clear dependence on the 
metallicity because collisions are less efficient at lower metallicities and 
thus NLTE deviations become larger. All this serves as a clear indication that 
NLTE effects have to be properly taken into account in Be abundance 
determinations.

Importantly, the Be abundance corrections are weakly 
sensitive to the effects of missing UV opacity. For example, the difference 
in the NLTE corrections for the Sun computed with and without taking the missing 
opacity into account is less than 0.015\,dex. This is because the 
corrections measure only the relative effect on the determined abundance, 
not its absolute value.
Clearly, the absolute 1D~NLTE abundance will be affected if the missing 
opacity is not taken into account.

\section{Conclusions}

We studied the influence of NLTE effects on the formation of \ion{Be}{ii} 
313.04 and 313.11\,nm resonance lines in solar and stellar atmospheres. 
For this, we constructed a new model atom of Be using the most up-to-date 
atomic data obtained in recent years using quantum mechanical 
computations. We find that NLTE effects play an important role both in the 
solar and stellar atmospheres and, depending on the combination of stellar 
parameters, metallicity, and Be abundance, may lead to either the strengthening or 
weakening of \ion{Be}{ii} lines. As a consequence, the solar Be abundance 
obtained in this study, ${\rm A(Be)_{\rm NLTE}}=1.32\pm0.05$, is 
slightly lower than the value obtained in earlier studies (cf. ${\rm A(Be)_{\rm NLTE}}=1.38\pm0.09$ from 
\citealt{Asplund21}) but is in excellent agreement with the 
meteoritic abundance determined by \citet{Lodders21}, 
${\rm A(Be)}=1.31\pm0.04$. We also provide a grid of NLTE--LTE 
abundance corrections to aid future Be abundance studies in late-type stars.

We also show that one of the important problems in Be abundance determinations 
is missing opacity in the blue--UV spectral range. No opacity 
source has been identified as being capable of making the synthetic and observed 
solar fluxes agree in this wavelength range. Clearly, work on improving 
opacities in this spectral range is of utmost importance, not only for the 
Be abundance studies but also for those of other elements whose lines are 
located at $250-390$\,nm.

\begin{acknowledgements}
We thank the referee Martin Asplund for his extensive and constructive comments that helped to improve the paper significantly. 
Our study has benefited from the European Union’s Horizon 2020 research and 
innovation programme under grant agreement No 101008324 (ChETEC-INFRA). This 
work has made use of the VALD database, operated at Uppsala University, the 
Institute of Astronomy RAS in Moscow, and the University of Vienna.
\end{acknowledgements}

%
%

\bibliographystyle{aa}
\bibliography{be}

\begin{appendix}

\section{List of spectral lines in the vicinity of \ion{Be}{ii} lines \label{app-sect:line-list}}

As described in Sect.~\ref{sect:modifications-line-list}, in order to determine Be abundance in the Sun 
we adjusted the parameters of several spectral lines located in the immediate neighbourhood of \ion{Be}{ii}  lines. 
The full line list in this spectral range is provided in Table~\ref{app-sect:line-list} and includes all lines 
that were used in spectral synthesis calculations for the determination of solar Be abundance (i.e. lines 
with the modified line parameters and lines with the original line parameters from the VALD database). 
The lines for which parameters were modified in accordance to the description provided in Sect.~\ref{sect:modifications-line-list} 
are marked in bold.

\begin{table*}[tb]
        \caption{Parameters of spectral lines in the vicinity of \ion{Be}{ii} lines.}
        \label{app-tab:line-list}
        \centering
        \begin{tabular}{cccc|cccc|cccc}
                \hline\hline            
                Species&Wave, nm&EP, eV&log gf&Species&Wave, nm&EP, eV&log gf&Species&Wave, nm&EP, eV&log gf\\
                \hline                      
                Ti 1& 312.96159&  1.4432& -2.108&Co 1& 313.03549&  2.8745& -2.781&OH 1& 313.09333&  0.6832& -3.360\\
                OH 1& 312.96363&  1.6950& -3.472&CN 1& 313.03659&  0.9898& -2.941&OH 1& 313.09926&  1.5694& -4.031\\
                Nb 2& 312.96522&  1.3209& -0.810&{\bf CH 1}&{\bf 313.03670}&{\bf 0.0350}&{\bf -1.828}&CN 1& 313.09959&  0.9476& -2.993\\
                Mn 1& 312.96523&  2.9197& -4.701&{\bf Ti 1}&{\bf 313.03772}&{\bf 1.4298}&{\bf -1.561}&Mn 2& 313.10150&  6.1113& -1.230\\
                OH 1& 312.96600&  1.7288& -2.439&Ce 2& 313.03835&  0.5210& -2.050&{\bf Mn 1}&{\bf 313.10161}&{\bf 3.7723}&{\bf -0.550}\\
                Mn 1& 312.96609&  3.7706& -1.377&CN 1& 313.03952&  0.9897& -2.958&CN 1& 313.10276&  0.9476& -3.011\\
                OH 1& 312.97078&  1.6950& -2.731&Be 2& 313.04200&  0.0000& -0.178&Be 2& 313.10650&  0.0000& -0.479\\
                Zr 2& 312.97600&  0.0390& -0.540&Mn 1& 313.04380&  3.7716& -2.699&Th 2& 313.10700&  0.0000& -1.559\\
                OH 1& 312.97619&  1.7288& -3.458&Ce 2& 313.04404&  0.8081& -1.950&Fe 2& 313.11041&  9.6878& -1.036\\
                Cr 1& 312.97701&  2.7079& -2.013&OH 1& 313.04694&  1.6089& -3.562&Fe 1& 313.11100&  3.0469& -5.610\\
                Fe 1& 312.97997&  2.8755& -4.043&OH 1& 313.04753&  1.9392& -2.042&Zr 1& 313.11100&  0.5203& -0.400\\
                CN 1& 312.98209&  1.0359& -2.908&Ce 2& 313.05137&  0.2323& -2.520&Os 1& 313.11160&  1.8409&  0.050\\
                Ru 1& 312.98410&  0.3850& -2.140&Cr 1& 313.05436&  4.1055& -1.337&Mo 1& 313.11940&  2.4992& -1.356\\
                Cr 1& 312.98669&  2.9674& -2.130&Mn 2& 313.05493&  6.4934& -1.120&CN 1& 313.11962&  0.3414& -4.225\\
                Cr 1& 312.99219&  3.9659& -3.207&Fe 2& 313.05611&  3.7677& -2.878&{\bf Cr 1}&{\bf 313.12156}&{\bf 3.1128}&{\bf -0.201}\\
                Y 2 & 312.99316&  3.4136&  0.753&OH 1& 313.05643&  2.1550& -3.881&CN 1& 313.12177&  0.3414& -4.251\\
                OH 1& 312.99380&  1.6089& -1.689&OH 1& 313.05694&  0.6832& -1.547&CN 1& 313.12184&  0.9345& -3.011\\
                Ta 1& 312.99430&  0.6969& -1.140&CN 1& 313.05696&  0.9753& -2.958&Fe 1& 313.12430&  2.1759& -3.792\\
                Mn 1& 312.99664&  4.3446& -1.208&Cr 2& 313.05716&  5.3297& -0.937&CN 1& 313.12506&  0.9345& -3.029\\
                Gd 2& 312.99680&  1.1719& -0.203&V 1 & 313.05748&  1.2181& -3.833&Tm 2& 313.12550&  0.0000&  0.240\\
                Th 2& 312.99740&  1.2868& -0.764&Ta 1& 313.05770&  1.3940&  0.070&Co 2& 313.13187&  2.2034& -3.801\\
                CN 1& 312.99787&  1.0201& -2.908&CN 1& 313.05998&  0.9752& -2.975&Mn 1& 313.13285&  4.6788& -1.746\\
                CN 1& 313.00057&  1.0200& -2.924&Cr 1& 313.06041&  3.5561& -2.550&Cu 1& 313.13326&  5.5057& -1.480\\
                Tm 2& 313.00484&  3.2211& -0.370&Mn 1& 313.06318&  4.2679& -0.946&Sc 2& 313.13387&  3.4223& -2.505\\
                Zr 1& 313.00560&  0.5190& -0.700&{\bf CH 1}&{\bf 313.06440}&{\bf 0.0350}&{\bf -1.520}&{\bf OH 1}&{\bf 313.13660}&{\bf 1.9420}&{\bf -1.150}\\
                CN 1& 313.00645&  0.3164& -4.306&Fe 1& 313.07044&  2.8755& -6.734&Fe 2& 313.13946&  3.8143& -3.784\\
                OH 1& 313.00746&  2.1550& -1.904&Mn 1& 313.07607&  4.2729& -2.903&OH 1& 313.14229&  0.9601& -2.363\\
                CN 1& 313.00832&  0.3163& -4.336&NH 1& 313.07701&  0.5610& -4.999&CN 1& 313.14471&  0.9218& -3.029\\
                OH 1& 313.01263&  0.8415& -2.207&CN 1& 313.07796&  0.9612& -2.975&Fe 1& 313.14590&  2.4688& -3.201\\
                Ti 1& 313.01671&  1.9807& -0.277&Nb 2& 313.07830&  0.4392&  0.410&CN 1& 313.14794&  0.9218& -3.048\\
                CN 1& 313.01691&  1.0047& -2.924&Rh 1& 313.07910&  0.4306& -2.110&Ce 2& 313.14980&  0.7345& -2.210\\
                Ce 2& 313.01932&  0.0000& -3.210&Ti 2& 313.07985&  0.0117& -1.190&OH 1& 313.15022&  0.4942& -2.695\\
                Fe 1& 313.01943&  3.5732& -3.253&Ce 2& 313.08044&  0.8085& -1.830&Fe 1& 313.15218&  3.2740& -3.730\\
                Ce 2& 313.01951&  0.2323& -2.140&CN 1& 313.08106&  0.9612& -2.993&Ni 1& 313.15298&  3.1930& -2.963\\
                CN 1& 313.01972&  1.0047& -2.941&Gd 2& 313.08130&  1.1566& -0.083&Cr 2& 313.15356&  4.1682& -1.548\\
                Mn 2& 313.02023&  4.8007& -3.556&OH 1& 313.08134&  1.9392& -3.783&NH 1& 313.15422&  0.5611& -4.866\\
                Fe 1& 313.02116&  3.3320& -2.196&Ce 2& 313.08517&  0.0000& -3.010&Hg 1& 313.15450&  4.8865& -0.040\\
                V 2 & 313.02701&  0.3483& -0.320&Ce 2& 313.08707&  1.0897& -2.250&Ce 2& 313.15457&  1.9299& -2.050\\
                OH 1& 313.02806&  0.2498& -1.964&Ce 2& 313.08757&  1.1069& -0.320&Cr 2& 313.15512&  4.1775& -1.602\\
                Ce 2& 313.03395&  0.5290& -0.590&Fe 2& 313.09046&  7.4867& -2.429&CN 1& 313.15752&  0.3507& -4.201\\
                Co 2& 313.03500&  2.9848& -3.459&Ce 2& 313.09217&  0.4954& -0.760&Mn 2& 313.15829&  6.4948& -1.892\\
                \hline                                                           
        \end{tabular}
\end{table*}

\section{1D~NLTE--LTE abundance corrections for the \ion{Be}{ii} 313.11\,nm line\label{app-sect:abund-corr}}

The 1D~NLTE--LTE abundance corrections for the \ion{Be}{ii} 313.11\,nm line are provided in Table~\ref{app-tab:NLTE_corrections}. 
Effective temperatures and surface gravities are given in Cols.~1 and 2, respectively, and the remaining
columns show abundance corrections for the \ion{Be}{ii} 313.11\,nm line computed for different [Fe/H] and [Be/Fe] ratios. 

\begin{table*}
        \caption{1D~NLTE--LTE abundance corrections for the \ion{Be}{ii} 313.11\,nm line.}
        \label{app-tab:NLTE_corrections}
        \centering
        \begin{tabular}{cc|ccc|ccc|ccc|ccc|ccc}
                \hline\hline            
                \multicolumn{2}{c}{ } & \multicolumn{3}{c}{[Fe/H]=-2.0}& \multicolumn{3}{c}{[Fe/H]=-1.0}& \multicolumn{3}{c}{[Fe/H]= 0.0}& \multicolumn{3}{c}{[Fe/H]=+0.5} \\
                \multicolumn{2}{c}{ } &\multicolumn{3}{c}{[Be/Fe]=}&\multicolumn{3}{c}{[Be/Fe]=}&\multicolumn{3}{c}{[Be/Fe]=}&\multicolumn{3}{c}{[Be/Fe]=}\\
                \hline                                   
                & &-0.5& 0.0&+0.5&-0.5& 0.0&+0.5&-0.5& 0.0&+0.5&-0.5& 0.0&+0.5\\
                \hline                                   
                \teff&\logg &\multicolumn{3}{c}{NLTE--LTE}&\multicolumn{3}{c}{NLTE--LTE}&\multicolumn{3}{c}{NLTE--LTE}&\multicolumn{3}{c}{NLTE--LTE}\\
                \hline                      
                4500 &  0.0  &+0.20 &+0.08&+0.02&-0.15&-0.19&-0.18&-0.23&-0.29&-0.31&-0.24&-0.31&-0.37\\
                4750 &  0.0  &+0.22 &+0.14&+0.12&-0.02&-0.03&+0.00&-0.24&-0.24&-0.19&-0.29&-0.34&-0.34\\
                5000 &  0.0  &+0.20 &+0.10&+0.06&+0.08&+0.08&+0.08&-0.13&-0.10&-0.03&-0.24&-0.22&-0.16\\
                5500 &  0.0  &+0.13 &+0.05&-0.04&+0.01&+0.00&-0.03&+0.00&+0.02&+0.09&-0.03&+0.00&+0.06\\
                6000 &  0.0  &+0.05 &+0.02&-0.09&-0.07&-0.12&-0.15&-0.05&-0.06&-0.02&-0.02&+0.00&+0.07\\
                4500 &  1.0  &+0.14 &+0.07&+0.00&-0.05&-0.09&-0.12&-0.12&-0.16&-0.22&-0.14&-0.18&-0.24\\
                4750 &  1.0  &+0.17 &+0.11&+0.07&-0.03&-0.05&-0.08&-0.16&-0.20&-0.22&-0.19&-0.25&-0.30\\
                5000 &  1.0  &+0.18 &+0.14&+0.08&+0.04&+0.04&+0.02&-0.14&-0.15&-0.13&-0.22&-0.26&-0.26\\
                5500 &  1.0  &+0.15 &+0.14&+0.06&+0.06&+0.04&+0.02&-0.02&-0.02&+0.01&-0.10&-0.09&-0.05\\
                6000 &  1.0  &+0.10 &+0.10&+0.02&+0.02&-0.04&-0.07&-0.02&-0.03&-0.03&-0.03&-0.03&+0.01\\
                6500 &  1.0  &+0.03 &+0.07&-0.02&-0.01&-0.07&-0.14&-0.07&-0.09&-0.11&-0.06&-0.07&-0.06\\
                4500 &  2.0  &-0.08 &-0.07&-0.10&-0.07&-0.11&-0.13&-0.05&-0.07&-0.11&-0.08&-0.08&-0.12\\
                4750 &  2.0  &-0.04 &-0.02&-0.05&-0.08&-0.12&-0.14&-0.07&-0.11&-0.15&-0.10&-0.13&-0.19\\
                5000 &  2.0  &+0.04 &+0.06&+0.02&-0.04&-0.07&-0.09&-0.09&-0.13&-0.15&-0.13&-0.18&-0.23\\
                5500 &  2.0  &+0.11 &+0.13&+0.09&+0.06&+0.03&+0.02&-0.05&-0.05&-0.06&-0.12&-0.14&-0.14\\
                6000 &  2.0  &+0.10 &+0.15&+0.08&+0.07&+0.03&-0.02&+0.00&-0.01&-0.03&-0.04&-0.05&-0.05\\
                6500 &  2.0  &+0.05 &+0.12&+0.03&+0.05&+0.02&-0.06&-0.02&-0.04&-0.07&-0.03&-0.05&-0.06\\
                4500 &  3.0  &-0.19 &-0.20&-0.19&-0.10&-0.11&-0.15&-0.02&-0.03&-0.05&-0.04&-0.02&-0.05\\
                4750 &  3.0  &-0.20 &-0.20&-0.18&-0.13&-0.14&-0.17&-0.03&-0.05&-0.08&-0.04&-0.04&-0.08\\
                5000 &  3.0  &-0.12 &-0.10&-0.10&-0.13&-0.14&-0.16&-0.05&-0.08&-0.11&-0.05&-0.08&-0.12\\
                5500 &  3.0  &+0.02 &+0.07&+0.03&+0.00&+0.00&-0.04&-0.06&-0.09&-0.11&-0.09&-0.13&-0.16\\
                6000 &  3.0  &+0.06 &+0.14&+0.06&+0.07&+0.07&+0.02&+0.00&-0.02&-0.04&-0.05&-0.07&-0.09\\
                6500 &  3.0  &+0.05 &+0.14&+0.04&+0.07&+0.08&+0.02&+0.02&-0.02&-0.05&-0.02&-0.03&-0.06\\
                4500 &  4.0  &-0.16 &-0.22&-0.13&-0.10&-0.10&-0.12&-0.02&-0.02&-0.04&-0.02&+0.01&-0.03\\
                4750 &  4.0  &-0.21 &-0.30&-0.20&-0.14&-0.13&-0.16&-0.03&-0.03&-0.06&-0.02&+0.00&-0.03\\
                5000 &  4.0  &-0.20 &-0.26&-0.19&-0.16&-0.16&-0.18&-0.04&-0.05&-0.08&-0.02&-0.02&-0.05\\
                5500 &  4.0  &-0.08 &-0.05&-0.06&-0.10&-0.08&-0.09&-0.06&-0.08&-0.12&-0.04&-0.07&-0.10\\
                6000 &  4.0  &+0.00 &+0.07&+0.00&+0.02&+0.05&+0.02&-0.03&-0.05&-0.08&-0.05&-0.08&-0.11\\
                6500 &  4.0  &+0.03 &+0.11&+0.01&+0.06&+0.10&+0.04&+0.03&+0.00&-0.04&-0.01&-0.04&-0.07\\
                4500 &  4.5  &-0.14 &-0.20&-0.10&-0.09&-0.09&-0.09&-0.03&-0.01&-0.05&-0.02&+0.01&-0.02\\
                4750 &  4.5  &-0.19 &-0.29&-0.18&-0.13&-0.14&-0.13&-0.04&-0.02&-0.06&-0.02&+0.00&-0.03\\
                5000 &  4.5  &-0.20 &-0.31&-0.20&-0.16&-0.17&-0.17&-0.04&-0.04&-0.07&-0.02&-0.01&-0.04\\
                5500 &  4.5  &-0.11 &-0.13&-0.08&-0.14&-0.13&-0.13&-0.07&-0.08&-0.11&-0.03&-0.04&-0.08\\
                6000 &  4.5  &-0.03 &+0.01&-0.02&-0.02&+0.02&-0.02&-0.05&-0.06&-0.10&-0.04&-0.07&-0.10\\
                6500 &  4.5  &+0.02 &+0.08&-0.01&+0.03&+0.09&+0.03&+0.01&+0.00&-0.04&-0.02&-0.04&-0.08\\
                4500 &  5.0  &-0.13 &-0.20&-0.09&-0.07&-0.07&-0.06&-0.04&-0.01&-0.04&-0.03&+0.01&-0.03\\
                4750 &  5.0  &-0.17 &-0.28&-0.15&-0.11&-0.14&-0.10&-0.04&-0.02&-0.06&-0.02&+0.01&-0.03\\
                5000 &  5.0  &-0.19 &-0.32&-0.20&-0.15&-0.19&-0.14&-0.05&-0.03&-0.07&-0.02&+0.00&-0.04\\
                5500 &  5.0  &-0.14 &-0.20&-0.10&-0.16&-0.18&-0.15&-0.07&-0.07&-0.11&-0.03&-0.03&-0.06\\
                6000 &  5.0  &-0.06 &-0.05&-0.03&-0.06&-0.03&-0.06&-0.07&-0.07&-0.11&-0.04&-0.06&-0.09\\
                6500 &  5.0  &-0.01 &+0.04&-0.02&+0.00&+0.06&+0.00&-0.01&-0.01&-0.05&-0.03&-0.05&-0.09\\
                \hline                  
        \end{tabular}
\end{table*}

\section{Comparison of NLTE--LTE abundance corrections for the \ion{Be}{ii} 313.11\,nm line with those from \citet{Takeda11}\label{app-sect:abund-corr-compar}}

One important finding of our study is that NLTE effects may strengthen the \ion{Be}{ii} resonance lines 
        significantly, both in the Sun and other types of stars (Sects.~\ref{sect:solar-Be-abundance} and \ref{sect:NLTE-corr}), 
        leading to significant NLTE--LTE abundance corrections (Sect.~\ref{app-sect:abund-corr}). This is in contrast to
        what was found in earlier studies of Be NLTE abundances in the Sun (Sect.~\ref{sect:solar-Be-abundance}). 
        As discussed in Sect.~\ref{sect:solar-Be-abundance}, we believe that the main reason for these differences is that 
        the Be model atom used in our study consists of a significantly larger number of levels and transitions and uses the most up-to-date 
        quantum mechanical data for treating the collisions with hydrogen (Sect.~\ref{sect:solar-Be-abundance}).

To assess how important these improvements could be for the realistic description of NLTE effects in Be line formation, 
        we used a simplified model atom consisting of ten levels of \ion{Be}{i}, five levels of \ion{Be}{ii,} 
        and the ground level of \ion{Be}{iii}, with the collisional rates with electrons and 
        hydrogen computed using the van Regemorter's and Drawin's formulae, respectively. 
        With these assumptions, the simplified model atom loosely matches the complexity of
        the model atom used in, for example, the study of \citet{Takeda11}.

The obtained abundance corrections are provided in 
        Table~\ref{app-tab:NLTE-corr-comparison}, along with those obtained in the study of \citet{Takeda11}, and
        show that there is a very good agreement between the corrections obtained in the two studies: for $\teff=5500$
        and $\log g=4.0$ the difference is only --0.01\,dex (ours versus theirs). Indeed, the differences between 
        the corrections become significant if the full Be model atom is used in our calculations: 
        for the same atmospheric parameters, we obtain the abundance correction of --0.08\,dex, 
        while \citet{Takeda11} predict zero abundance correction.

\begin{table*}
        \caption{Comparison of the NLTE--LTE abundance corrections for the \ion{Be}{ii} 313.11\,nm line obtained in this study with those determined by \cite{Takeda11}.}
        \label{app-tab:NLTE-corr-comparison}
        \centering
        \begin{tabular}{ccccccccccc}
                \hline\hline            
                Teff &log g& [Fe/H]&\vmicro& A(Be)&  NLTE--LTE (T11) & NLTE--LTE (this work)\\
                \hline                      
                5000 &4.0&   0.00 & 1.5&  1.15&  +0.01&       0.00\\
                5500 &4.0&   0.00 & 1.5&  1.15&   0.00&      -0.01\\
                6000 &4.0&   0.00 & 1.5&  1.15&  +0.02&      +0.02\\
                6500 &4.0&   0.00 & 1.5&  1.15&  +0.09&      +0.07\\
                \hline                      
        \end{tabular}
\end{table*}

\section{Be abundance corrections for the most metal-poor stars\label{app-sect:abund-corr-3}}

In their recent study of Be abundances in the most metal-poor Galactic stars, Smiljanic et al. (2021) suggested 
that at the lowest metallicity end the Be abundances may shown indications for the existence of  `Be-plato'. 
Since the Be abundance corrections obtained in our study do show a tendency to increase with decreasing 
metallicity, one may therefore wonder how large NLTE effects may be in the most metal-poor Galactic stars. 
We therefore computed Be abundance corrections for the \ion{Be}{ii} 313.11\,nm line for stars with various 
combinations of \teff\ and $\log g$ at $\feh=-3.0$. The results are provided in Table~\ref{app-tab:abund-corr--3}
for cases where the equivalent width of the \ion{Be}{ii} 313.11\,nm line is ${\rm EW}>1$\,pm.

Clearly, for some combinations of stellar parameters and Be abundances, the abundance corrections 
may become significant. However, even in the most extreme cases this would not be sufficient to account 
for the $\sim1$\,dex excess in Be abundances obtained in the LTE analysis of Smiljanic et al. at $\feh\leq-3.0$.

\begin{table*}
        \caption{1D~NLTE--LTE abundance corrections for the \ion{Be}{ii} 313.11\,nm line 
                at [Fe/H]= --3.0 (for cases with EW $>1$\,pm). }
        \label{app-tab:abund-corr--3}
        \centering
        \begin{tabular}{cc|ccccc}
                \hline\hline            
                \multicolumn{2}{c}{[Be/Fe]=}&-0.5& 0.0&+0.5&+1.0&+1.5\\
                \hline                                   
                \teff&\logg &\multicolumn{5}{c}{NLTE--LTE}\\
                \hline                      
                4500 &  0.0  &+0.60&+0.39&+0.24&+0.05&-0.05\\
                4750 &  0.0  &+0.52&+0.36&+0.25&+0.07&-0.02\\
                5000 &  0.0  &+0.43&+0.32&+0.24&+0.06&-0.04\\
                5500 &  0.0  &     &+0.24&+0.14&+0.03&-0.12\\
                6000 &  0.0  &     &     &+0.03&+0.00&-0.15\\
                4500 &  1.0  &+0.38&+0.30&+0.29&+0.11&+0.04\\
                4750 &  1.0  &+0.33&+0.28&+0.26&+0.11&+0.04\\
                5000 &  1.0  &+0.28&+0.27&+0.25&+0.13&+0.02\\
                5500 &  1.0  &     &+0.25&+0.18&+0.12&+0.01\\
                6000 &  1.0  &     &     &+0.07&+0.09&-0.02\\
                6500 &  1.0  &     &     &     &+0.05&-0.05\\
                4500 &  2.0  &     &+0.08&+0.01&+0.00&-0.08\\
                4750 &  2.0  &     &+0.10&+0.03&+0.02&-0.06\\
                5000 &  2.0  &     &+0.15&+0.07&+0.06&-0.01\\
                5500 &  2.0  &     &     &+0.10&+0.09&+0.06\\
                6000 &  2.0  &     &+0.17&+0.16&+0.13&+0.06\\
                6500 &  2.0  &     &+0.12&+0.11&+0.10&+0.02\\
                4500 &  3.0  &     &     &-0.15&-0.15&-0.17\\
                4750 &  3.0  &     &     &-0.13&-0.14&-0.15\\
                5000 &  3.0  &     &     &-0.08&-0.07&-0.10\\
                5500 &  3.0  &     &     &+0.04&+0.05&+0.01\\
                6000 &  3.0  &     &     &     &+0.13&+0.05\\
                6500 &  3.0  &     &     &     &+0.13&+0.03\\
                4500 &  4.0  &     &     &     &-0.20&-0.14\\
                4750 &  4.0  &     &     &     &-0.24&-0.18\\
                5000 &  4.0  &     &     &     &-0.21&-0.17\\
                5500 &  4.0  &     &     &     &-0.06&-0.07\\
                6000 &  4.0  &     &     &     &+0.03&-0.01\\
                6500 &  4.0  &     &     &     &     &+0.01\\
                4500 &  4.5  &     &     &     &     &-0.12\\
                4750 &  4.5  &     &     &     &     &-0.16\\
                5000 &  4.5  &     &     &     &-0.26&-0.16\\
                5500 &  4.5  &     &     &     &-0.14&-0.09\\
                6000 &  4.5  &     &     &     &     &-0.03\\
                6500 &  4.5  &     &     &     &     &-0.01\\
                4500 &  5.0  &     &     &     &     &     \\
                4750 &  5.0  &     &     &     &     &-0.15\\
                5000 &  5.0  &     &     &     &     &-0.16\\
                5500 &  5.0  &     &     &     &     &-0.10\\
                6000 &  5.0  &     &     &     &     &-0.04\\
                \hline                  
        \end{tabular}
\end{table*}

\end{appendix}
\end{document}